\documentclass{iau}
\usepackage{graphicx}

\title[Pre-main sequence star magnetic fields] 
{The role of magnetic fields in pre-main sequence stars}

\author[Gaitee A.J. Hussain \& Evelyne Alecian]   
{Gaitee A.J. Hussain$^1$ \& Evelyne Alecian$^{2,3}$}

\affiliation{$^1$ESO, Karl-Schwarzschild-Strasse 2, D-85748, Garching bei M\"unchen \\ Germany \\
email: {\tt ghussain@eso.org} \\[\affilskip]
$^2$UJF-Grenoble 1 / CNRS-INSU, Institut de Plan\'etologie et d'Astrophysique de Grenoble (IPAG) UMR 5274, Grenoble, F-38041, France \\email: {\tt evelyne.alecian@obs.ujf-grenoble.fr}\\[\affilskip]
$^3$LESIA, UMR 8109 du CNRS, Observatoire de Paris, UPMC, Universit\'e Paris Diderot, 5 place Jules Janssen, F-92195 Meudon Cedex, France}
\pubyear{2013}
\volume{302}  
\pagerange{119--126}
\jname{IAUS 302: Magnetic fields throughout stellar evolution}
\editors{M. Jardine, P. Petit \& H. Spruit, eds.}
\begin{document}

\newcommand{\msun}{\,\mbox{$\mbox{M}_{\odot}$}}

\maketitle

\begin{abstract}

Strong, kilo-Gauss, magnetic fields are required to explain a range of observational properties in young, accreting pre-main sequence (PMS) systems. 
We review the techniques used to detect magnetic fields in PMS stars. Key  results from a long running campaign aimed at characterising the large scale magnetic fields  in accreting T Tauri stars are presented. Maps of surface magnetic flux in these systems can be  used to build 3-D models exploring the role of magnetic fields and the efficiency with which magnetic fields can channel accretion from circumstellar disks on to young stars. Long-term variability in T Tauri star magnetic fields strongly point to a dynamo origin of the magnetic fields. Studies are underway to quantify how changes in magnetic fields affect their accretion properties. We also present the first results from a new programme that investigates the evolution of magnetic fields in intermediate mass (1.5--3\msun) pre-main sequence stars as they evolve from being convective  T Tauri stars to fully radiative Herbig AeBe stars.

\end{abstract}

\firstsection 
\begin{figure}[ht]
\includegraphics[width=5.5in]{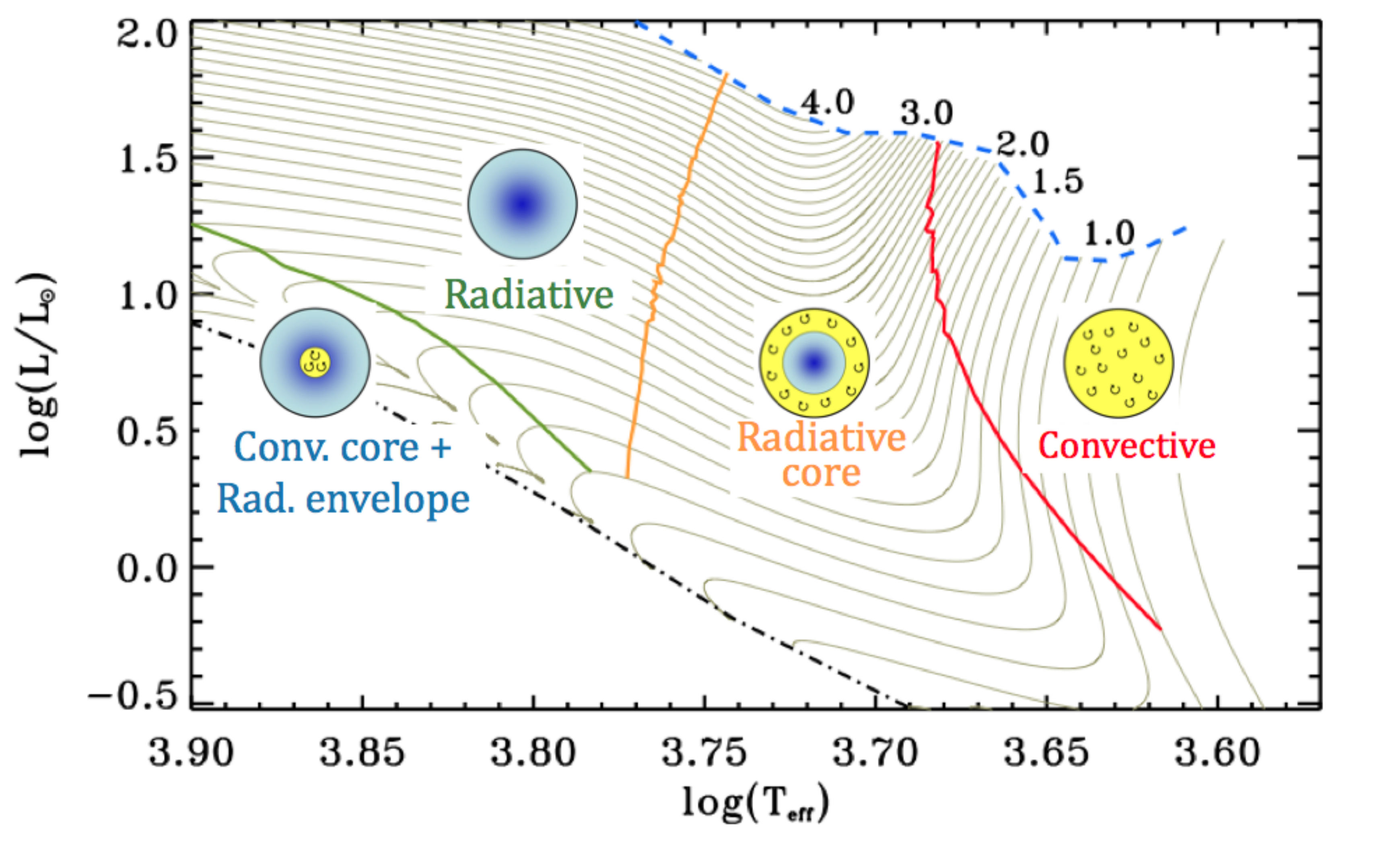} 
\caption{H-R diagram showing Behrend \& Maeder (2001)  pre-main sequence evolutionary tracks for stellar masses up to 4\msun. The dashed blue line marks the position of the birthline. 
All stars with masses less than 3.5\msun\ will undergo a stage along their pre-main sequence evolution in which they have either partially or fully convective interiors.  A star with a mass of 1.5\msun\ or more will be subject to  several fundamental changes in their internal structure, having a fully convective interior near the birthline, to developing a radiative core, to becoming fully radiative and finally developing a convective core just before reaching the Zero Age Main Sequence (black dot-dashed line).}
   \label{figevo}
\end{figure}

\section{Background}

Pre-main sequence (PMS) models show that all stars form from a fully convective low mass core. In Fig.\,\ref{figevo}, stars rise along the birth line (blue dashed line) until the strong accretion phase stops and then follow a quasi-static contraction along PMS tracks. We adopt Behrend \& Maeder (2001) tracks here; different models predict slightly different positions for the PMS tracks (Siess et al. 2000, Tognelli et al. 2011) but they all show the same fundamental transitions to internal structure with mass. 
Higher mass stars naturally evolve much more quickly along these tracks compared to their lower mass counterparts. 
It is worth noting that A-type stars will undergo fundamental changes to their interior structures during their PMS evolution;
starting as fully convective T Tauri stars, developing radiative cores; eventually becoming fully radiative Herbig stars
(Grunhut et al., {\em this volume}, Alecian et al. 2013). Finally, just before they reach the Zero Age Main Sequence they will  develop
 a convective core.
 This review focusses on the magnetic properties of T Tauri stars, which either have fully convective interiors or  outer convective envelopes. While these stars cover a relatively narrow range in spectral type (G to M), they encompass almost two orders of magnitude in stellar luminosity.  

It has long been a requirement for T Tauri stars to host kilo-Gauss magnetic fields with strong dipoles in order to explain several key observational characteristics.
Classical T Tauri stars, T Tauri stars that are accreting,  have long rotation periods  (typically between 7--10d); these are much longer than predicted from angular momentum conservation considering the stars are contracting and actively accreting material from their circumstellar disks. 
Observations indicate  discrete regions  at which accretion streams impact the stellar surface; shocks form near the photosphere that can be detected at  X-ray and UV wavelengths. These XUV diagnostics indicate material impacting at near-free fall velocities at or near the photosphere. The model that can best explanation most of the observational properties of classical T Tauri stars is the magnetospheric accretion model. In this model,  a strong dipole field from the star extends several stellar radii, truncating the inner edge of the stellar disk and channelling disk material along these magnetic field lines in streams that then impact the stellar surface. From a study of the classical T Tauri star, AA Tau, 
Bouvier et al. (2007) find that both the photometric variability and the modulation seen in its Balmer line can be explained in terms of accretion funnels that pass along the line of sight and a magnetically warped disk that  periodically occults the star. 

The magnetospheric accretion environment can regulate the angular momentum of these young systems through accretion powered stellar winds, ejections through reconnection events and similar interactions between the stellar magnetosphere and the disk (e.g., Hartmann \& Macgregor 1982; Matt \& Pudritz 2005, 2008, Zanni \& Ferreira 2013). Once accretion stops or becomes less efficient then stars are free to spin up as they contract towards the Main Sequence.

\section{PMS star magnetic fields: detection techniques}
\label{sec:techniques}

Stellar magnetic fields can be directly detected and characterised using two main techniques, {\em Zeeman broadening}  and {\em Zeeman Doppler imaging}. Both of these are spectroscopic techniques that require high resolution spectra ($R>30\,000$) and rely on the Zeeman effect. In the presence of a magnetic field atomic and molecular lines can show  broadening or even full splitting, depending on the magnetic sensitivity of the line  and the size of the magnetic field. Lines with no magnetic sensitivity are magnetic null lines and are useful diagnostics of the non-magnetic photosphere.
A brief description of these techniques as applied to T Tauri stars is given here, along with a list of the main advantages and disadvantages associated with each (Table\,\ref{tab1}). The interested  reader is referred to  the review by Donati \& Landstreet (2009) for a more detailed description  of the Zeeman effect and these techniques.

\subsection{Zeeman broadening}

The Zeeman broadening technique measures the broadening in intensity line profiles with different magnetic sensitivities ($g$, Land\'e-factors). 
It has been used to measure magnetic field strength distributions and magnetic fluxes in a range of T Tauri stars (e.g., Johns-Krull 2007, 2008, Yang et al. 2011 \& references therein).  The size of the broadening scales with the square of the wavelength (Eqn.\,\ref{eq1}), so longer wavelengths yield more robust measurements and measurements of the mean  magnetic field. The surface magnetic fields of T Tauri stars have been measured using a set of Ti I lines near 2.2$\mu$ that have different magnetic sensitivities. 
Magnetic null ($g_{\rm eff}=0$) CO lines near 2.3$\mu$  are used to characterise non-magnetic photospheric parameters, e.g., veiling ,Teff, vsini, microturbulence.
This technique is particularly effective when applied to T Tauri stars, as many are relatively slow rotators and have spectral types of K or later. Ti I 2.2$\mu$ lines are best suited to K-M spectral types as they weaken in hotter stars.

\begin{equation}
\Delta \lambda_B = 4.67 \lambda^2_{\circ}g_{\rm eff}B .
\label{eq1}
\end{equation}

The  wavelength broadening,  $\Delta \lambda_B$, scales with the stellar magnetic field, $B$, the wavelength of the line $\lambda_{\circ}$, and its effective Land\'e factor, $g_{\rm eff}$, a measure of the mean magnetic sensitivity of the line. 
Zeeman Broadening studies have shown that multiple magnetic field strength components are often required to fit the observed splitting in T Tauri stars if applied to spectra encompassing lines with different magnetic sensitivities. 
A mean magnetic field strength can be computed from these different components as follows: $ {Bf}=\Sigma B_i f_i$, with a typical range of field strengths ($B_i$) between  2\,kG to 6\,kG, and associated filling factors ($f_i$) for each component ranging between 20-50\%.

\subsubsection{Zeeman Broadening: strengths and challenges}
\label{seczbchallenges}
The  main strengths and challenges associated with  this technique are summarised in the left column of Table\,\ref{tab1} and expanded on in this section. 
A key strength is that it uses the information in Stokes I intensity profiles, hence the magnetic field measurements are sensitive to the strongest surface magnetic fields  observable on the projected stellar disk, even if these are  concentrated in complex small scale active regions. This technique is not subject to flux cancellation unlike  techniques which utilise circularly polarised signatures (Stokes V profiles).

 The modelling in Zeeman Broadening measurements assumes a purely radial field (e.g., Johns-Krull 2007, Yang et al. 2011). While extreme orientations could potentially be discerned (e.g., fields entirely aligned parallel to -- or perpendicular to -- the line-of-sight),  very little information on the field topology can be obtained from these spectra alone. Yang et al. (2011) confirm previous findings that the Ti~I spectra of T Tauri stars show little evidence for extreme orientations such as these. If the surface fields are composed of a mixture of  radial, azimuthal (East-West) and meridional (North-South) orientations, the magnetic field measurements will be affected  as different field orientations alter the strengths of the $\pi$ and $\sigma$ components of each line, and therefore the shapes of the magnetically sensitive line profiles. In the absence of further information radial field orientations are, however, the simplest assumption.

Zeeman Broadening measurements are best applied to low $v_e \sin i$ stars, typically less than 20\,km/s.
Significant Doppler broadening makes it difficult to measure the Zeeman broadening, $\Delta \lambda_B$ effectively.
It should be noted that it has however been applied successfully up to $v_e \sin i \sim 55$\,km/s (the M1.5 T Tauri star, TWA 5a; Yang et al. 2008). 
Further assumptions used in Zeeman broadening  measurements are that the magnetic field regions  are uniformly distributed over the stellar surface; clearly if the magnetic field regions were concentrated at the poles the required filling factors would change. 
The temperature structure associated with the magnetic field regions is also assumed to be uniform, i.e., not preferentially concentrated in cool spotted regions (Johns-Krull 2007). As these studies are usually based on single-epoch spectra, no information on the latitudinal positions or inhomogeneity of the surface fields can be obtained. Finally, the multi-component solutions are non-unique: multi-component fields are not always required to achieve similar levels of agreement with the data (Yang et al. 2011). This degeneracy can introduce a 10--15\% uncertainty in the mean magnetic field strength measurements. 

\begin{table}[h]
  \begin{center}
  \caption{Magnetic field measurement techniques: pros \& cons (Sec.\,\ref{seczbchallenges} \& \ref{seczdichallenges})}
  \label{tab1}
 {\scriptsize
  \begin{tabular}{l|c|cc}\hline
 &  {\bf Zeeman Broadening } & {\bf Zeeman Doppler Imaging   }\\ 
\hline
{\bf Pros}   &  Measure strongest magnetic fields 	&  Recovers large-scale field topology \\ 
           	& Insensitive to field geometry 			&  Photospheric \& accretion diagnostics\\
\\
{\bf  Cons} & Snapshot single-epoch {\bf \em Bf} measurements 
										& Flux cancellation (circularly polarised spectra) \\
                    &  Assume uniform temperature 		
                    								& Phase coverage \& S:N affect magnetic flux strength\\
              	& 2.2$\mu$ studies limited to K-M type stars 
										& Limited dark spot information  \\
              	& Slow rotation ($v_{e}\sin i < 30$km/s) & Missing field information from inclined hemisphere\\
             	&  Non-unique solution of multi-component {\bf \em B} fields 
										& Non-unique solution (regularising functions) \\
   
 \hline
  \end{tabular}
  }
 \end{center}
\end{table}

\subsection{Zeeman Doppler Imaging}
 Doppler imaging techniques have been used to invert time-series of intensity spectra to recover surface maps of inhomogeneities, including brightness distributions, temperatures, abundances in a range of stars  from Ap stars to cool M dwarfs (e.g., Vogt, Penrod \& Hatzes 1987; Piskunov, Tuominen \& Vilhu 1990; Barnes \& Collier Cameron 2001). 
Zeeman Doppler imaging techniques apply Doppler imaging principles to circularly polarised (Stokes V) signatures in order to reconstruct the large scale magnetic fields on the surfaces of magnetically active stars (Semel 1989, Donati \& Cameron 1997).
Zeeman Doppler imaging exploits a key characteristic  of Stokes~V profiles, that they are predominantly sensitive to the line-of-sight component (longitudinal component) of the magnetic field. In cool stars these signatures are very weak, typically at 0.1\% of the continuum level. A single-epoch Stokes V spectra enables us to robustly detect the  presence of the longitudinal magnetic field. As demonstrated by Donati et al. (1997), the longitudinal component of $B$, $B_l$ at a particular epoch can be estimated from  the first moment of the Stokes V profile. 

A time-series of Stokes V  profiles, covering a full stellar rotation period, enables us to track the modulation of the line-of-sight component of the stellar magnetic field caused by magnetic regions crossing the projected stellar disk. This rotational modulation enables us to pinpoint not only the location of these regions in latitude and longitude, but also their relative field orientations. 

When applied to T Tauri stars that are still accreting this technique can be applied simultaneously to both the photospheric and accretion line profiles  (e.g., He I D$_3$ at 5876\AA\ or the near-infrared Ca II triplet at 8500\AA). This is possible thanks to the development of large-format high resolution \'echelle spectrographs\footnote{The primary facilities optimised for these studies are CFHT/ESPaDOnS, TBL/NARVAL and the ESO 3.6-m/HARPS.} that cover over a thousand photospheric lines and several useful accretion diagnostics in T Tauri stars. 
It is necessary to sum up the signature from hundreds of photospheric lines using the cross-correlation technique, Least Squares Deconvolution (LSD)  in order to detect stellar magnetic field signatures robustly (Donati et al. 1997, Kochukhov et al. 2010, Chen \& Johns-Krull 2013). This is because the sizes of the signatures are typically small, at 0.1\% of the continuum level, even in magnetically active stars.
 
  In  classical T Tauri stars  strong emission lines often show significant circular polarisation in individual lines. In the majority of systems observed the Ca II NIR triplet shows strong Stokes V signatures indicating kG fields associated with the accretion. In some systems where the Ca II NIR Stokes V signatures are found to be weak (e.g., V4046 Sag), they support the picture of  a complex large scale stellar magnetic field; with flux cancellation in Stokes V being caused by accretion spots with multiple polarities.

\subsubsection{ZDI: Strengths \& Challenges}
\label{seczdichallenges}

The main limitations and challenges associated with  this technique are summarised in Table\,\ref{tab1} and expanded on in this section.  The spatial resolution of the magnetic field maps obtained with Zeeman Doppler Imaging depends on the $v_e \sin i$ of the star, the spectroscopic resolution and the phase coverage obtained. In low $v_e \sin i$ stars ($<20$km/s), while it is possible to discriminate between simple and complex topologies ,flux cancellation results in an inability to measure the strongest fields at the stellar surface. Hussain et al. (2009) investigate whether it is possible to discriminate between simple dipole-dominant fields and more complex fields in low  $v_e \sin i$ stars and find that even though the spatial resolution is more limited at low  $v_e \sin i$'s, more complex large scale fields will be detected using Zeeman Doppler imaging.

Large gaps in phase coverage can result in smearing of magnetic field regions as their exact positions become harder to pinpoint. This also weakens the size of the magnetic flux size recovered in the map. The magnetic field maps obtained of T Tauri stars have been obtained using Doppler imaging codes that assume a two-component brightness model, consisting of an immaculate photosphere and a cool, dark spotted component. As spots are dark the relative flux contribution is limited and so it is inevitable that magnetic fields concentrated in the darkest spotted regions cannot be detected in photospheric line profiles. What is notable is that, in common with active G and K-type stars on the Main Sequence (e.g., Donati \& Cameron 1997),  a significant fraction of what appears to be the unspotted ``immaculate'' photosphere is found to be strongly magnetic. This may imply that the whole stellar surface is spotted at small scales that cannot currently be recovered using  Doppler imaging techniques. 

Finally, as with all Doppler imaging techniques, the solutions obtained are non-unique. It is possible to fit an observed time-series of Stokes I and V spectra with many different magnetic field and brightness distributions within a specified level of $\chi^2$ agreement. It is therefore necessary to employ a regularising function, which enables a unique robust solution to be obtained. The maps presented here use Maximum Entropy, which minimises the amount of information needed to fit the observed spectroscopic time-series. Hussain et al. (2000) present a comparison of images obtained with two independent Maximum Entropy Zeeman Doppler imaging codes and find the images to be very similar though the exact form of the entropy used determines the sharpness of the structure reconstructed in the surface maps.

\section{Zeeman Broadening measurements of {\bf \em B} and {\bf \em Bf} }

\begin{figure}[ht]
\includegraphics[width=5.in]{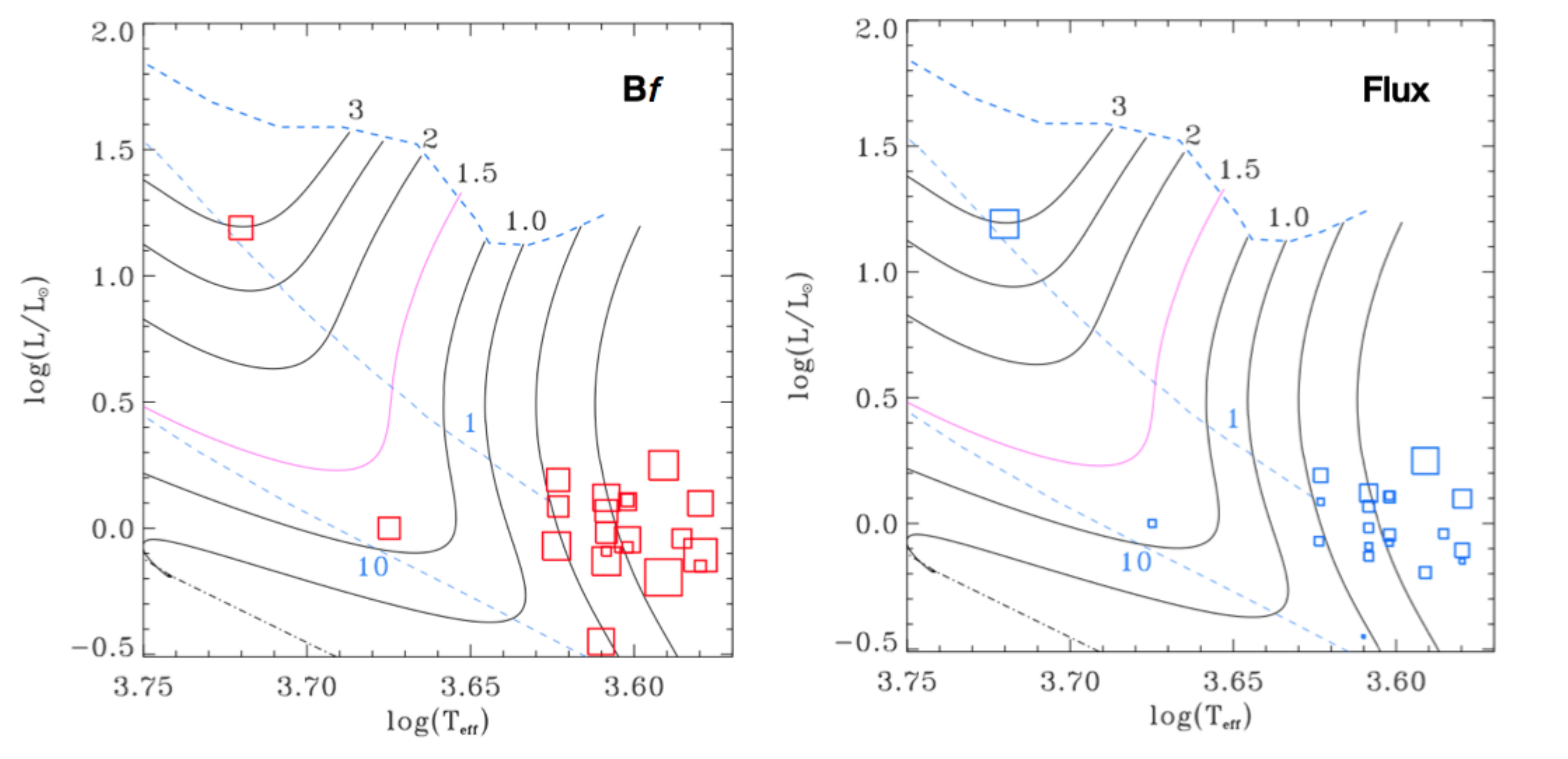} 
\caption{Surface magnetic fields in  T Tauri stars measured using the Zeeman broadening technique:  the symbol sizes scale with mean magnetic field strength ({\em left} ) and mean magnetic flux ({\em right}); $Bf$-values are from Yang et al. (2011) and references therein. Behrend \& Maeder (2001) evolutionary tracks are over-plotted for PMS stars with masses between 0.6 and 3\msun.}
   \label{figbf}
\end{figure}

Over 30 T Tauri stars have been studied using the Zeeman Broadening technique, including several classical T Tauri stars. Figure\,\ref{figbf} shows the resulting mean magnetic field strengths, $Bf$ (left) and the mean magnetic fluxes plotted in an H-R diagram. These plots also show evolutionary tracks for PMS stars with masses ranging from 0.7 to 3\msun, using the Behrend \& Maeder (2001) tracks. All published measurements are shown here, excluding the lowest mass T Tauri stars with masses below 0.55\msun. 

In the left-hand plot of Fig.\ref{figbf}, the symbol sizes reflect the  mean magnetic field strengths, $Bf$, ranging from 1.1 to 3.5\,kG, with 1\,kG being the lower limit of the measurements. No clear trend with age or stellar mass is visible in the sizes of the mean $Bf$ value. Indeed the highest and lowest average $Bf$ are both found  in stars with similar M0-1 spectral types: DE Tau and LO Ori, which have mean magnetic field strengths of 1.12\,kG and 3.45kG respectively. The prototype of the class, T Tau, is the highest mass star in the sample shown here ($M_*\sim 3$\msun\ according to the Behrend \& Maeder tracks) and clearly  has a similar average  $Bf$  to that found in  its lower mass counterparts. 

The large magnetic field strengths recovered are typically much larger than those required by pressure equipartition considerations, $B_{eq} = (8\pi P_g)^{1/2}$; where $P_g$ is the gas pressure at the atmospheric height corresponding to the observed line formation (Johns-Krull 2007). As T Tauri stars have low surface gravities and  relatively low gas pressures, flux tube equilibrium models predict lower magnetic fields than are found from Zeeman Broadening measurements. This implies that no equilibrium can exist between magnetic and ``non-magnetic'' regions (if they exist) in T Tauri stars. Caution must therefore be exercised in interpreting activity phenomena on these stars as simple scaled up versions of those observed on the Sun and solar-type stars. 

The magnetic field strengths measured from these techniques are also generally significantly stronger than those predicted  by simple analytic models of magnetospheric accretion (Johns-Krull 2007, 2008). 
These models estimate the  sizes of the surface dipolar fields required given the radius, mass and accretion rate of a particular T Tauri star.  Hence it is likely that even though T Tauri stars host adequately strong fields to channel accretion from circumstellar disks,  the fields are unlikely to  be described in terms of a simple dipole field aligned with the rotation axis in the majority of the observed systems.

On the right-hand plot of Fig.\,\ref{figbf} , the symbol sizes scale with the mean flux for each star, $F_B=4\pi R^2\bar B$; where $\bar B$ is the mean magnetic field strength and $R$ is the stellar radius. As T Tauri stars age and contract towards the main sequence their changing stellar sizes drive the corresponding decrease in the  mean magnetic flux.
A possible interpretation is that these stars are moving from fossil-type simple unchanging fields to more complex fields with a dynamo origin as they approach the Zero Age Main sequence (Yang et al. 2011).
On the other hand, temporal changes seen in magnetic field maps of accreting T Tauri stars obtained from Zeeman Doppler imaging techniques over a period of years imply a dynamo-type mechanism and appear to argue against a simple ``frozen-in'' fossil field origin in the youngest T Tauri stars
(Sec.\,\ref{zditemp}).

\section{Zeeman Doppler Imaging: magnetic field maps}

\begin{figure}[ht]
\includegraphics[width=5.5in]{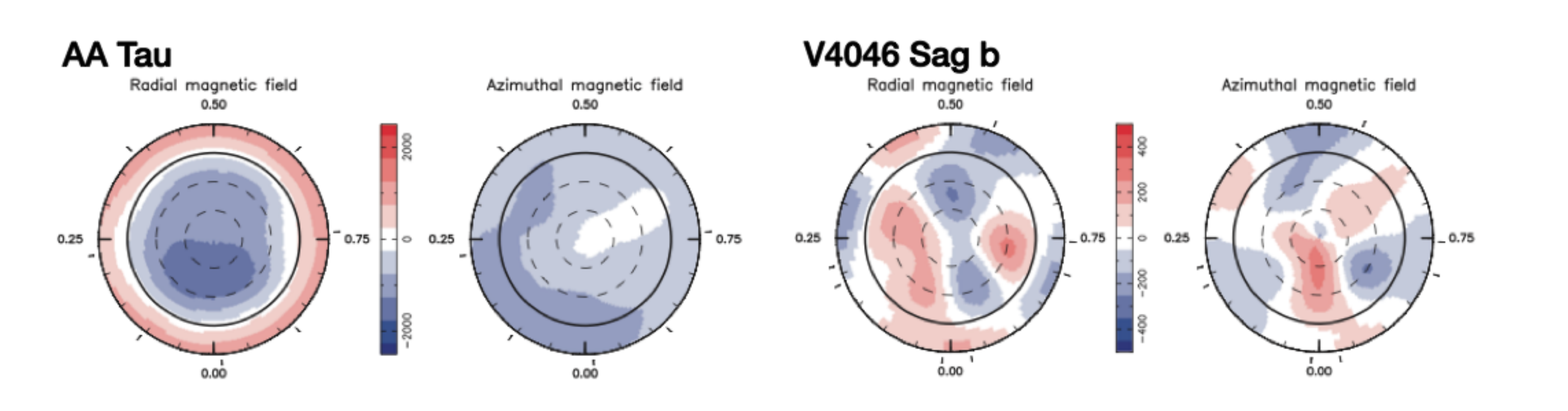} 
\caption{Radial and azimuthal field maps for two classical T Tauri stars; AA Tau ({\em left}) and V4046 Sag b ({\em right}). 
V4046 Sag b, has a significantly more complex field than AA Tau despite the similar spatial resolution of the maps. Magnetic flux maps obtained to date show that  fully convective stars possess relatively simple strong AA Tau-type fields, while  
stars with large radiative cores display complex fields. These maps are polar projections of the stellar surface, with concentric rings marking rings of 30 degree latitudes down to -30 degrees (the equator is the thick solid circle). The colour scales represent magnetic fluxes in G - note the different scales in the two sets of maps. These maps are reproduced from Donati et al. (2010, 2011).}
   \label{figzdmaps}
\end{figure}

The surface magnetic field maps of twelve classical T Tauri stars have been reconstructed using Zeeman Doppler imaging techniques, mostly under the framework of an international campaign led by Jean-Francois Donati entitled, ``Magnetic Protostars and Planets''  (MaPP\footnote{Further information at: http://lamwws.oamp.fr/magics/mapp/MappScience.}). As these stars are actively accreting it is possible to study their surface magnetic fields and their accretion properties simultaneously using the same dataset, which contains thousands of photospheric lines as well as several accretion-sensitive diagnostics such as He~I, Ca II H\&K, several Balmer lines and the Ca II NIR triplet. The list of stars for which maps have been produced and published include:  V2129 Oph, BP Tau, V2247 Oph, AA Tau, TW Hya, V4046 Sag a \& b, GQ Lup, DN Tau, CR Cha, CV Cha, and MT Ori (Donati et al. 2007, 2008b, 2010a,b, 2011a,b,c, 2012, 2013; Hussain et al 2009; Skelly et al. submitted).

While relatively few T Tauri star systems have been studied some clear trends  are emerging. 
Their large scale fields can be characterised as lying between two extreme cases as shown in Fig. \,\ref{figzdmaps}.
This figure illustrates the radial field and azimuthal (east-west) oriented field maps for two classical T Tauri stars, AA Tau and V4046 Sag b. While they have different masses they have similar basic parameters (AA Tau: $v_e \sin i=11.3$\,km/s \& $T_{\rm eff}=4000$\,K;  V4046 Sag b: $v_e\sin i=13.5$\,km/s \& $T_{\rm eff}=4250$\,K).  As these stars have such similar $v_e \sin i$'s and the data used have the same spectroscopic resolution (from CFHT/ESPaDOnS) both sets of  maps have comparable spatial resolution scales.

It is immediately apparent from Fig.\,\ref{figzdmaps} that V4046 Sag b has a significantly more complex field distribution than AA Tau, this is reflected in the multiple switches in polarity observed in both the radial and azimuthal surface field maps. AA Tau maps also show a stronger magnetic flux ($\pm 2$kG) compared to V4046 Sag b  ($\pm 0.4$kG). While this may indicate that V4046 Sag b has a weaker surface magnetic field, V4046 Sag b maps are more likely to be affected by flux cancellation due to the star's  more complex large scale field. 

\begin{figure}[ht]
\includegraphics[width=5.5in]{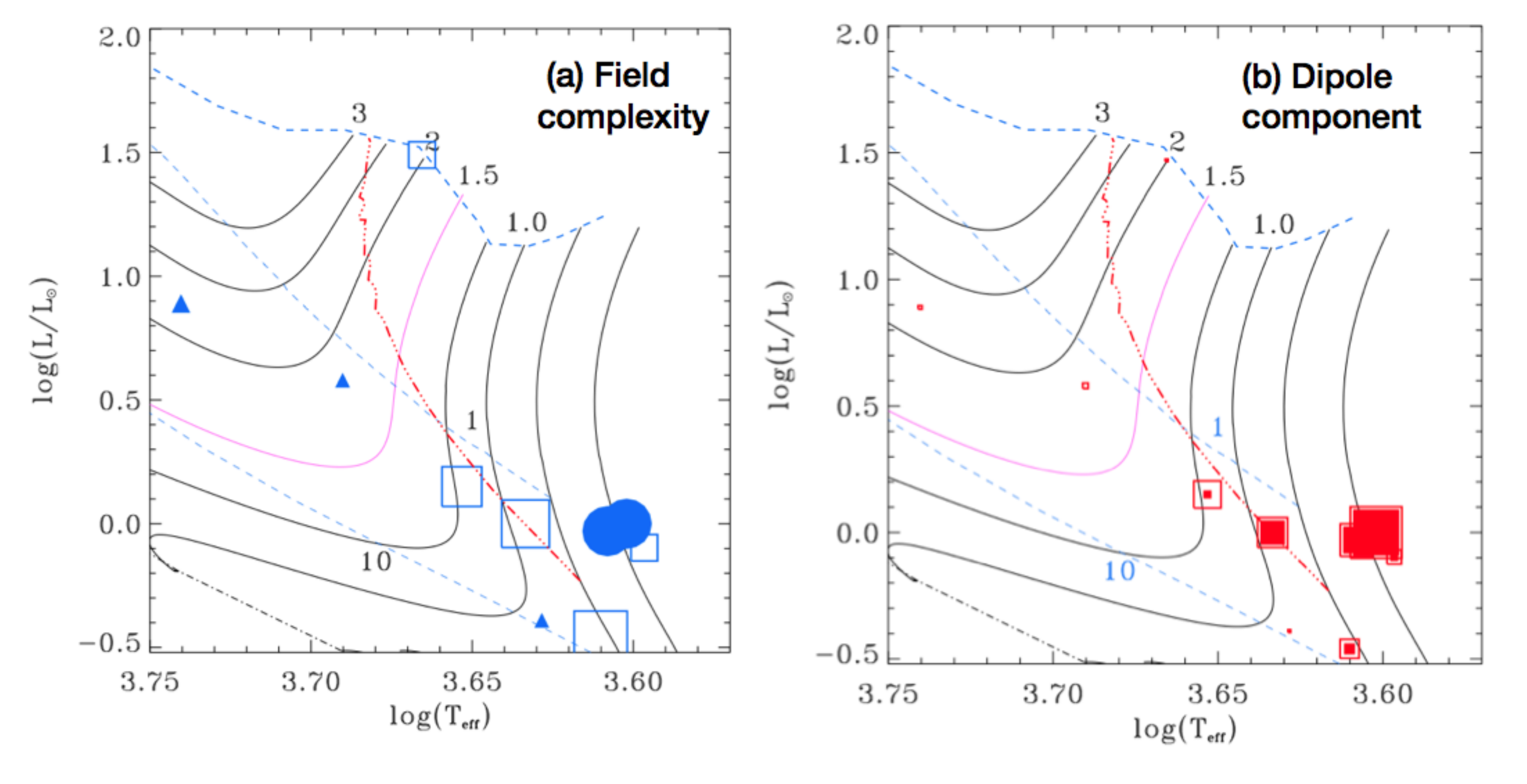}
\caption{Surface magnetic fields in  T Tauri stars from Zeeman Doppler imaging. {\em Left}: Symbol sizes scale with the magnetic intensity. Different symbols denote the dominant field orientation: circles -- dipole-dominant fields; squares -- octupole-dominant fields; triangles -- complex multipolar fields. {\em Right}: Symbols scale with the strength of the dipole field component (Gregory et al. 2012); where multi-epoch observations were taken the stronger dipole field measurement is represented by an unfilled square for clarity. The largest change in the dipole field is found in maps of V2129 Oph taken 4 years apart.
 Behrend \& Maeder (2001) evolutionary tracks are over-plotted as in Fig.\,\ref{figbf}. Stars develop a radiative core to the left of the dot-dashed line.}
   \label{figzdbevo}
\end{figure}

 Fig.\,\ref{figzdbevo} shows a plot summarising key properties of the magnetic field maps obtained for all of the classical  T Tauri stars maps that have been studied to date.  The sizes of the symbols scale with the magnetic intensities in Fig.\,\ref{figzdbevo}a, from  0.4-0.5\,kG  (in CR Cha and V4046 Sag a\,\&\,b) to 4\,kG (in GQ Lup). 
 The dominant mode of the large-scale field is represented by the symbol shape; circles show dipole-dominant fields, open squares are octupole-dominant fields and  triangles are for maps with higher degrees of complexity. The dot-dashed line marks the division between fully convective  stars and stars with  radiative cores. It is clear that fully convective stars have the strongest simplest fields; the closer they get to the dividing  line they become octupole-dominant; and as the radiative cores grow the  fields become more complex.

Fully convective stars with masses lower than 0.7\msun\ may have more complex fields: DN Tau (0.65\msun)
 is octupole-dominant even though fully convective (Donati et al. 2013). The fully convective star, V2247 Oph, has an even more complex field; it  is not pictured in Fig.\,\ref{figzdbevo} as it has a very low mass (0.35\msun).  The reason for the increased complexity on the lowest mass T Tauri stars is discussed in Sec.\,\ref{conclusions}.

Four stars show complex fields: CV Cha, CR Cha, V4046 Sag a and b (Hussain et al. 2009, Donati et al. 2011). These stars may cover a range of masses from 0.9 to 2.5\msun\ but they all have large radiative cores ($M_{\rm core}\gtrsim 0.4M_*$).  Their Stokes V signatures show complex structure and rotational modulation indicating the complex field maps recovered from Zeeman Doppler imaging; accretion diagnostics provide further support. The Ca~II NIR Stokes V profiles of V4046 Sag a and b show weak polarisation despite having  a similar mass accretion rate to AA Tau, TW Hya and GQ Lup. This may be caused by flux cancellation in multiple accretion streams with opposite polarities. No Stokes V signatures were detected in He~I and Balmer line profiles of CV Cha and CR Cha. CV Cha has  the highest mass accretion rate of all the stars studied to date ($\log \dot M$$\sim$$-7.5$\msun\,yr$^{-1}$), a simple large scale field would easily have been detected in its strong emission lines. 

Symbol sizes scale with the dipole field strengths in  Fig.\,\ref{figzdbevo}b (Gregory et al. 2012). As found in various studies, this property is particularly important in determining the disk truncation radius (e.g., Johnstone et al. 2013). It is clear that as complexity increases the dipole field strength drops, which should have a corresponding effect on the accretion state of the star.

\subsection{Temporal evolution}
\label{zditemp}
Zeeman Doppler imaging maps have been acquired at multiple epochs for five T Tauri star systems, BP Tau, AA Tau, V2129 Oph, GQ Lup and DN Tau (Donati et  al. 2008,2010b, 2012, 2013.  All of the stars studied so far have shown changes in their large scale fields. Despite these changes the global properties of the fields of  a particular star remain similar ( i.e., a star with an octupole-dominant field does not become dipole-dominant).

\begin{figure}[ht]
\includegraphics[width=5.5in]{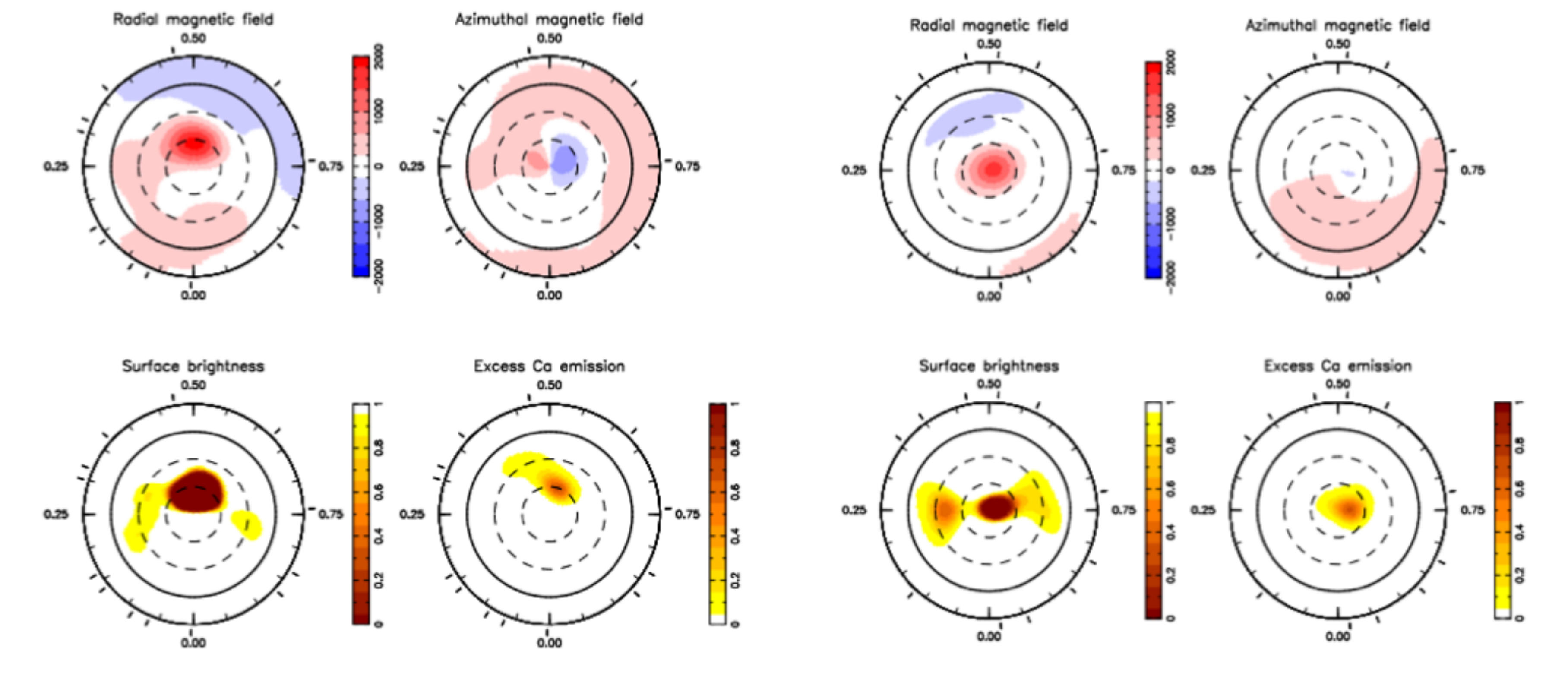}
\caption{Temporal evolution in the large scale field of DN Tau. Radial and azimuthal field maps, brightness maps and excess Ca II emission maps of DN Tau acquired two years apart (Left: Dec 2010 \& Right: Dec 2012). As in Fig.\,\ref{figzdmaps} these are polar projections of the stellar surface and the observed phases are denoted by tick marks around each map. While the field strength recovered is similar at both epochs the octupole:dipole field strength  appears to increase in 2012 and  may cause the changing shape and position of the accretion spot, as traced by excess Ca II emission.  Figures from Donati et al. (2013).}
 \label{figzdtemporal}
\end{figure}

DN Tau maps acquired two years apart show that there are  changes in its large scale magnetic field as well as changes in  the shape of the accretion spot (Fig.\,\ref{figzdtemporal}, Donati et al. 2013). The accretion spot appears more circular and polar in 2012; at the same time the magnetic field maps show that the relative strength of the octupole to dipole field component has increased. The changing shape of the accretion region from a crescent shape to a circular shape accompanied by a suggestion of an increasingly dominant octupole field is in agreement with predictions from 3-D MHD simulations of disc accretion (e.g., Romanova et al. 2011). However, an effort is needed to quantify how accurately the relative strengths of the different field components can be measured in ZDI maps.  

Models can use these magnetic field maps as inputs to predict the locations of  accretion spots; these can be compared directly with the observed accretion maps (e.g., the Ca II emission maps in Fig.\,\ref{figzdtemporal}). 
Observations of GQ Lup taken at three separate epochs over a period of two years suggest that the field can change significantly within one year (Johns-Krull et al. 2013, Donati et al. 2012). Further monitoring of the large scale fields of these stars is planned in the framework of a new large programme, MaTYSSE (see Sec.\,\ref{conclusions}, this will enable us to monitor and characterise the variability in both the accretion and stellar magnetic fields in T Tauri stars  over a 10-year period.

\section{Summary \& Conclusions}
\label{conclusions}
\begin{itemize}
\item Mean magnetic field strengths measured on 33 T Tauri stars range between 1-3.5\,kG. While 
there are no clear trends in the mean $Bf$ values with stellar parameter; the mean magnetic fluxes clearly 
decrease with age (by $\sim$30\% in the first Myr); this is predominantly due to the shrinking radii of the evolving PMS stars.

\item The magnetic field strength measurements indicate the presence of very strong fields, with strengths up to 6\,kG,  at the surfaces of several T Tauri stars. Comparisons with analytic models suggest that T Tauri star magnetic fields are likely not organised in simple dipoles in most cases. 

\item The large scale fields have been mapped on twelve accreting T Tauri stars as part of a larger campaign, MaPP, 
which aims to study the large scale magnetic fields in T Tauri stars and their influence on their accretion states. 

\item MaPP results indicate that stars with fully convective
 envelopes have large scale axisymmetric dipole dominant fields, these change to octupole-dominated fields once  a radiative core develops, with even more complex large scale fields  found  in stars with large radiative cores ($M_{\rm core}\gtrsim0.4M_*$).  This dependence of the field on the internal structure has clear analogies with the magnetic field studies of main sequence M dwarfs, where fully convective M dwarfs tend to possess simpler large scale fields than their higher mass counterparts,  which have a radiative core (Morin et al. 2011).
\item The picture may change again in  lower mass T Tauri stars ($M_*\lesssim 0.6$\msun) which fall into a bistable dynamo regime and may have more complex fields(Gregory et al. 2012). Of the 4 fully convective stars studied, AA Tau and BP Tau ($M_*\sim 0.7-0.75$\msun)  are nearer the fully convective limit and show dipole-dominant fields. In contrast, the lower mass stars, DN Tau (0.65\msun) and V2247 Oph (0.35\msun), are deeper in the fully convective phase and show increasingly more complex fields that are similar to the lowest mass M dwarfs (Donati et al. 2013).

\item The large scale fields of classical T Tauri stars show temporal changes over a period of years. MHD accretion models predict that changing field topologies have an impact on the shape and distribution on the accretion spots. 
More observations  will enable us to characterise and quantify the changes in the large scale field and accretion over a period of ten years. At the same time 3D MHD models can use these maps to evaluate the impact of these changing fields on the accretion properties of the stars in detail. 
\end{itemize}

The coming years will bring many more advances in our understanding of magnetism in PMS stars. In particular the recently started large programmes on the CFHT, 
MaTySSE\footnote{{\em MaTYSSE:} Magnetic Topologies of Young Stars \& the Survival of close-in massive Exoplanets (PI: Donati).} and  BinaMIcS\footnote{ {\em BinaMIcS:} Binarity and Magnetic Interactions in various classes of Stars (PI: Alecian).} will contribute to this. The former will reconstruct the large scale magnetic fields in T Tauri stars that have stopped accreting to investigate how stellar magnetic fields and  dynamos are affected by this change in accretion state.
BinaMIcS will investigate  magnetospheric interaction in close T Tauri binary star systems.

\begin{figure}[ht]
\includegraphics[width=5.5in]{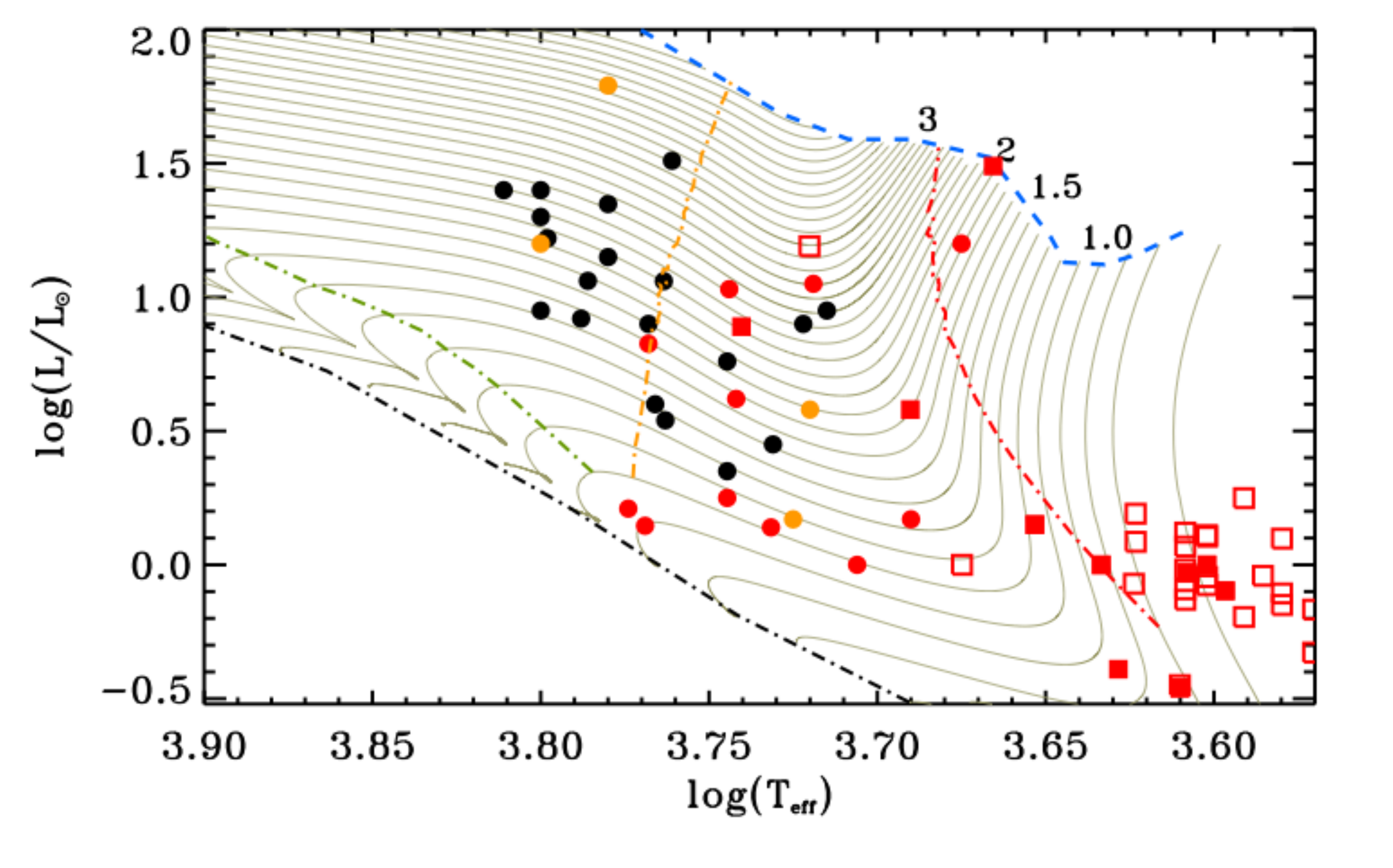}
\caption{Going to higher mass: The incidence of magnetic fields in T Tauri stars across a range of masses. Evolutionary tracks are solid lines and the dot-dashed lines show where stars develop radiative cores (red line), become fully radiative (orange) and develop convective cores (green; also see Fig.\,1).
 T Tauri stars for which magnetic fields have been detected are marked in red (open squares -- Zeeman broadening measurements; filled squares -- Zeeman Doppler imaging; circles -- spectro-polarimetric observations described below). Marginal detections are denoted by orange circles.}
 \label{figbfimtts}
\end{figure}

\subsection{Intermediate mass T Tauri stars}
\label{secimtts}

From  a spectro-polarimetric study analysing homogeneous data acquired of 70 Herbig Ae/Be stars
Alecian et al. (2013) confirm the detection of five magnetic stars. 
They find a low incidence (10\%) of stellar magnetic fields in Herbig AeBe systems. This  is comparable to  that seen in main sequence A and B-type stars:  with between 5--10\% of A and B-type stars host strong magnetic fields (Donati \& Landstreet 2009). The properties of the magnetic Herbig Ae/Be stars are also found to be similar to those seen on their main sequence counterparts  (Alecian et al. 2013, also see Grunhut et al., {\em this volume}). We have  recently started a study to characterise the magnetic field properties of the precursors to Herbig Ae stars, the intermediate mass T Tauri stars ($1.5<M_*<3$\,\msun).

We are probing the magnetic field incidence for over 40 intermediate mass T Tauri stars, which have been identified from their effective temperatures and luminosities in the literature. Magnetic fields are detected using snapshot Stokes V spectra obtained from the CFHT and ESO 3.6-m telescopes. For stars for which we have acquired multi-epoch Stokes V spectra, we can begin to  analyse the  large scale field topology. The ultimate goal is to characterise the magnetic field in intermediate mass T Tauri stars as they evolve into fully radiative Herbig stars (Alecian, Hussain et al. in prep.). 

Fig.\,\ref{figbfimtts} shows the first results from our programme;  the intermediate mass T Tauri stars are filled circles. On this figure we over-plot the T Tauri stars for which magnetic fields have been measured and analysed using the Zeeman broadening and Zeeman Doppler imaging techniques. Dot-dashed lines denote where PMS stars interiors undergo key changes.  It is clear that the incidence of stellar magnetic fields drops off with increasing $T_{\rm eff}$; i.e., as the stars lose their convective envelopes. A detailed spectroscopic analysis is underway to  obtain  better constraints on the fundamental properties of each of the target stars and to place strong upper limits on any non-detections. However, these first results support the idea that the internal structure of the star, not  mass, is a key determinant of stellar dynamo behaviour -- at least in G--M-type PMS  stars.

\subsection{Multi-wavelength studies}
Multi-wavelength campaigns are particularly effective ways to test details of magnetospheric accretion models. Argiroffi et al. (2010) study the X-ray emission in  the close binary T Tauri system, V4046 Sag, as part of a coordinated XMM-{\em Newton}/CFHT campaign. They find that the soft high density X-ray emission in the system  shows significant orbital modulation, while the hotter plasma associated with the stellar corona shows no evidence of  orbital modulation. This modulation is best explained in terms of high density plasma  concentrated near the stellar photosphere in compact regions that are eclipsed as the system rotates. Fonseca et al. show the power of combining multi-wavelength photometry with Balmer line spectroscopy  to investigate the relationship between the structure of the inner accretion disk and stellar magnetosphere.

Further detailed tests of magnetospheric accretion models will be possible by using large scale field  maps of T Tauri stars as inputs to  3-D MHD models. These models are being used to investigate how different large scale field geometries affect the accretion geometries in systems and can be tested against observations through detailed modelling of  profiles that form in the accretion funnels, e.g., H$\beta$ and He~I~10830\AA, 
(e.g., Kurosawa et al. 2012, Adams \& Gregory 2012).

\end{document}